\documentclass[pra,aps,showpacs,superscriptaddress,twocolumn]{revtex4-2}
\usepackage{graphicx,amsmath,amssymb,color}%,subfig}
\usepackage{xcolor}
\usepackage[colorlinks, citecolor={blue!50!black}, urlcolor={blue!50!black}, linkcolor={red!50!black}]{hyperref}
\begin{document}

\title{Entanglement enhancement in cavity magnomechanics by an optical parametric amplifier}

\author{Bakht Hussain}
\affiliation{Department of Physics and Applied Mathematics, Pakistan Institute of Engineering and Applied Sciences (PIEAS), Nilore $45650$, Islamabad, Pakistan.}
\author{Shahid Qamar}
\affiliation{Department of Physics and Applied Mathematics, Pakistan Institute of Engineering and Applied Sciences (PIEAS), Nilore $45650$, Islamabad, Pakistan.}
\affiliation{Center for Mathematical Sciences, PIEAS, Nilore, Islamabad $45650$, Pakistan.}
\author{Muhammad Irfan}
\affiliation{Department of Physics and Applied Mathematics, Pakistan Institute of Engineering and Applied Sciences (PIEAS), Nilore $45650$, Islamabad, Pakistan.}
\affiliation{Center for Mathematical Sciences, PIEAS, Nilore, Islamabad $45650$, Pakistan.}

\date{\today}

\begin{abstract}

We propose a method to enhance bipartite and tripartite entanglement in cavity magnomechanics using an optical parametric amplifier (OPA).
We analyze this system and identified parametric regimes where different types of entanglement are enhanced. 
We show that, a proper choice of phase of the parametric amplifier leads to the enhancement of the bipartite entanglements.
Moreover, the tripartite entanglement is also significantly enhanced in the presence of OPA.
The OPA not only enhances the strength of entanglement but also increases the domain of entanglement over a wider space of detunings as compared to the system when no OPA is present.
Similarly, the robustness of entanglement against temperature is also enhanced.
Another important consequence of OPA is the fact that it relaxes the requirement of strong magnon-phonon coupling to generate cavity-magnon entanglement which is necessary for the case when OPA is not present.
We believe that the presented scheme is a step forward to realize robust quantum entanglement using current technology.

\end{abstract}
\maketitle
\newpage
\section{Introduction}
Cavity magnomechanics, where a ferrimagnetic crystal (e.g., yttrium iron garnet (YIG)) is coupled with a microwave cavity has drawn considerable attention during last few years.
It provides a unique platform to study interactions among cavity, magnon, and phonon modes.
Magnetic materials generally have high spin density and low dissipation rate.
This results in strong~\cite{huebl2013high,tabuchi2014hybridizing,zhang2014strongly,goryachev2014high,bai2015spin,zhang2015cavity} and ultrastrong~\cite{bourhill2016ultrahigh,kostylev2016superstrong} couplings among the magnon or Kittel mode~\cite{kittel1948theory} (collective excitation of the spins) and a microwave cavity mode.
This strong coupling enables the coherent information transfer between different excitations which has a wide range of applications in quantum information processing.

Macroscopic quantum entanglement plays an important role not only in quantum information processing but also to understand fundamental theories.
In this regard, cavity optomechanics attracted a lot of attention during past decade~\cite{florianRMP}.
Vitali et al. studied the existence of quantum entanglement between the cavity mirror and the cavity field~\cite{vitali2007optomechanical}.
Recent experiments showed promising results where entanglement between cavity-phonon~\cite{palomaki2013, riedinger2016} and two mechanical oscillators~\cite{ockeloen-korppi2018, riedinger2018} is demonstrated.
Such macroscopic entangled states are proposed to help understand fundamental issues of modern physics like quantum-gravity~\cite{marletto2017}, quantum-to-classical transition~\cite{frowis2018}, testing models of wave-function collapse~\cite{bassi2013}, etc.
On applications side, such macroscopic quantum systems allow precision measurement and sensing, better quantum memory, and single photon detection~\cite{florianRMP, barzanjeh2021}.

Recently, in a seminal work, Li et al.,~\cite{li2018magnon} showed that similar to optomechanics, cavity magnomechanics also offers possibility to generate macroscopic entanglement.
This system supports both bipartite and tripartite entanglement.
Subsequent studies discussed entanglement between two magnon modes~\cite{li2019entangling, yu2020macroscopic, nair2020deterministic}, two vibrational modes of YIG spheres~\cite{li2021entangling}, steady Bell state generation~\cite{yuan2020}, among others.
Later, it is shown that entanglement can be further enhanced by including Kerr effect~\cite{zhang2019quantum, yang2020entanglement}.
In a very recent study, it is shown that the presence of injected weak squeezed vacuum field~\cite{PhysRevA.99.021801} enhances entanglement and one-way steering, especially in the red-detuned driving regime~\cite{Zhang-22}.
Similarly, it is shown that the resonant interaction of the antiferromagnet with light enhances the magnon-magnon entanglement~\cite{PhysRevB.101.014419}.

In this study, we consider the effects of a degenerate optical parametric amplifier on entanglement properties of the cavity magnomechanical system.
We show that the presence of the optical parametric amplifier (OPA) enhances the bipartite as well as tripartite entanglement for properly chosen parameters.
Apart from enhancement, it also widens the parameter space where entanglement can exist.
Therefore, makes the system more feasible towards experimental realization. % by imposing less stringent conditions on parameter space. 
Moreover, entanglement in the presence of OPA is more robust against the thermal excitations.
Li et al., showed that the bipartite entanglement between cavity and magnon modes exist only for a non-zero value of magnon-phonon coupling~\cite{li2018magnon} and maximum entanglement is achieved when the coupling is sufficiently strong.
We found that remarkably, this condition is no longer needed in the presence of OPA and strong cavity-magnon entanglement exists even for zero magnon-phonon coupling.
\begin{figure}[tbh] 
\includegraphics[width=1.\linewidth]{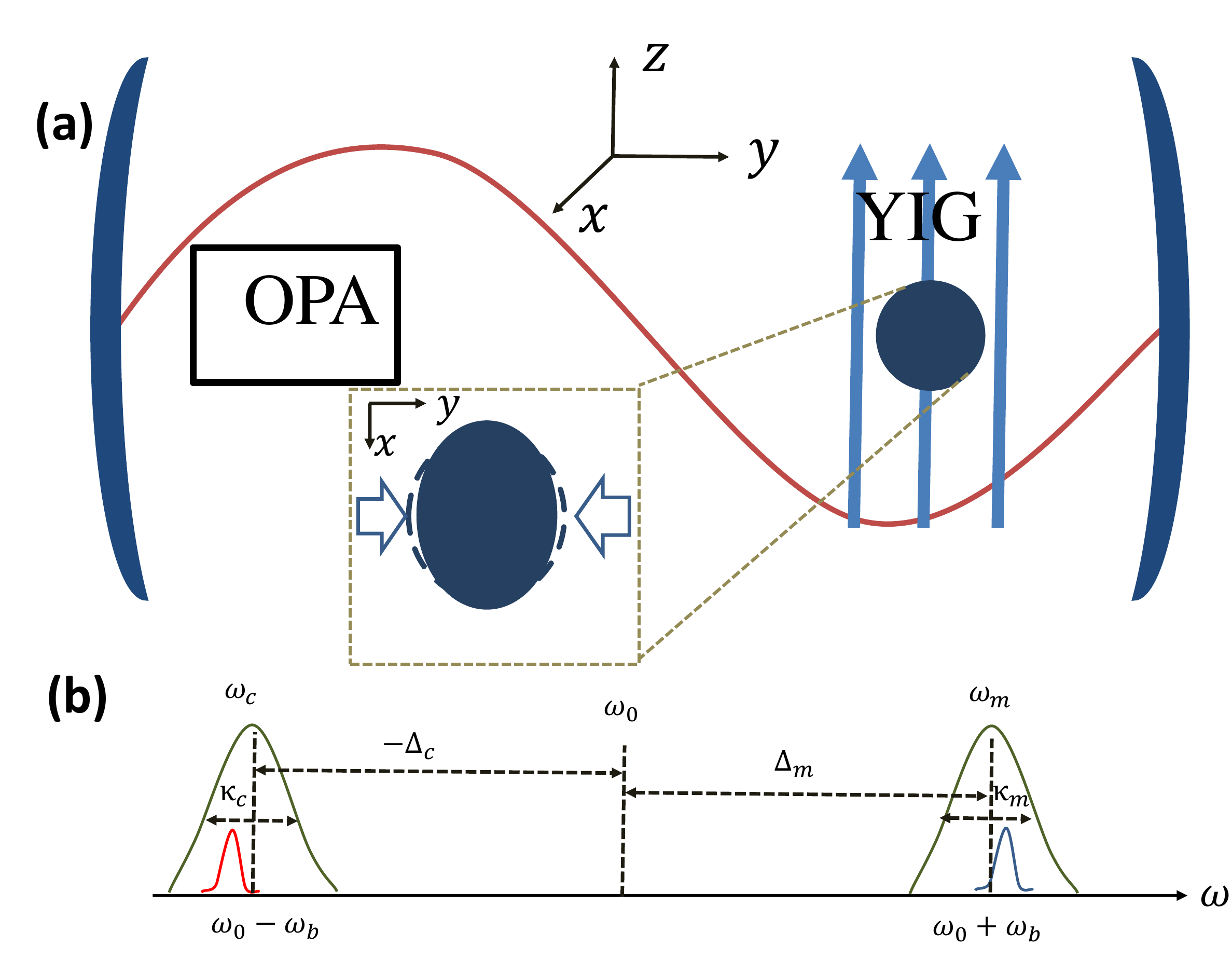}
\caption{\label{fig:1}
(a) Schematic diagram of a single-mode cavity with OPA and a YIG sphere. YIG sphere is placed in cavity such that it is simultaneously near the maximum magnetic field of the cavity mode and in a uniform bias field, which is responsible for magnon-photon coupling. A microwave field (not shown) is applied to enhance magnon-phonon coupling. At YIG sphere site, the magnetic field (along $x$ axis) of the cavity mode, the drive magnetic field (in $y$ direction), and bias magnetic field ($z$ direction) are mutually perpendicular. (b) System frequencies and linewidths.  A microwave field at frequency $\omega_0$ drives the magnon mode $m$ with frequency $\omega_m$ and bandwidth $\kappa_m$, and the photons scatter by mechanical motion at frequency $\omega_b$ onto the two sidebands at $\omega_0\pm\omega_b$.}
\end{figure}
 \section{system model and Hamiltonian}
 We consider a hybrid cavity-magnomechanical system~\cite{zhang2016cavity} which consists of a single-mode cavity of frequency $\omega_c$ with an OPA and a YIG sphere as shown in Fig.~\ref{fig:1}.
 This system supports three different types of excitation namely, photons, magnons, and phonons.
 The magnon mode is due to the collective motion of a large number of spins in a ferrimagnetic system e.g., a YIG sphere with a typical diameter of $250\,{\mu}m$ \cite{zhang2016cavity}.
 The coupling between the cavity mode and magnon mode is mediated by the magnetic dipole.
 The magnetostrictive interaction leads to the coupling of magnons and phonons.
 The phonon mode arises due to the geometric deformation of YIG sphere.
 The deformation is caused because of the magnon excitation inside the YIG sphere which induces a varying magnetization~\cite{kittel1958interaction}.

The Hamiltonian of the system under rotating-wave approximation in a frame rotating with the frequency of the drive field is given by:
 \begin{align}\label{E1}
 H/\hbar=&\Delta_c c^{\dag}c+\Delta_m m^{\dag}m+\frac{\omega_b}{2}(q^2+p^2)+g_{mb}m^{\dag}mq \\
 &+g_{mc}(c^{\dag}m+cm^{\dag})\nonumber+i G(e^{i\theta}c^{\dag2}-e^{-i\theta}c^2)\\
 &+i\Omega(m^{\dag} - m).\nonumber
\end{align}
The first two terms in Eq.~(\ref{E1}) describe the energy of the bare cavity and magnon modes, respectively with annihilation operators $c$ and $m$ and creation operators $c^\dag$ and $m^\dag$.
The detuning of the cavity frequency $\omega_c$ from drive field is $\Delta_c=\omega_c-\omega_0$, while of magnon mode frequency $\omega_m$ is $\Delta_m=\omega_m-\omega_0$.
The third term represents the mechanical mode of frequency $\omega_b$ (phonon mode), where $q$ and $p$ are dimensionless position  and momentum quadratures.
The interaction between the magnon and phonon modes is characterized by the magnomechanical coupling rate $g_{mb}$ which is typically small.
However, applying a strong drive microwave field enhances this rate.
Magnon and microwave cavity coupling is described by the fifth term with magnon-cavity coupling rate $g_{mc}$.
The condition for strong coupling is: $g_{mc}>\kappa_c,\kappa_m$~\cite{huebl2013high,tabuchi2014hybridizing,zhang2014strongly,goryachev2014high,bai2015spin,zhang2015cavity}, where $\kappa_c$ and $\kappa_m$ are dissipation rates of cavity and magnon modes, respectively.
The parameter $G$ represents the nonlinear gain of the OPA with $\theta$ the phase of the driving field.
A microwave field of amplitude $B_0$ and frequency $\omega_0$ drives the magnon mode.
The strength of the drive is characterized by the Rabi frequency $\Omega=\frac{\sqrt{5}}{4}\gamma\sqrt{N}B_0$~\cite{li2018magnon}, where $\gamma/2\pi=28~GHz/T$ is gyromagnetic ratio and $N=\rho{V}$ is the total number of spins with YIG sphere spin density $\rho=4.22\times10^{27}m^{-3}$ having volume $V$.
The derivation of $\Omega$ assumes low-lying excitations i.e.,  ${\langle}m^{\dag}m\rangle{\ll}2Ns$, where ground state spin number of $Fe^{3+}$ ion in YIG is $s=\frac{5}{2}$.
Here, we neglect the effects of the radiation pressure because the wavelength of the microwave field is assumed to be much larger than the size of the sphere.
Similarly, the rotating-wave approximation is justified under the condition: $\omega_c$, $\omega_m \gg g_{mc}$, $\kappa_c$, $\kappa_m$~\cite{zhang2016cavity}.
The quantum Langevin equations (QLEs)~\cite{benguria1981quantum} describing the system are given by:
\begin{align}\label{E2}
\dot{c}=&-(i\Delta_c+\kappa_c)c-ig_{mc}m+2Ge^{i\theta}c^\dag+\sqrt{2\kappa_c}c^{in},\nonumber\\
\dot{m}=&-(i\Delta_m+\kappa_m)m-i g_{mc}c-i g_{mb}mq+\Omega+\sqrt{2\kappa_m}m^{in},\nonumber\\
\dot{q}=&~\omega_bp, \nonumber \\
\dot{p}=&-\omega_bq-\gamma_bp-g_{mb}m^{\dag}m+\xi,
\end{align}
 with $\gamma_b$ the mechanical damping rate, whereas $c^{in}$, $m^{in}$ and $\xi$ are input noise operators for cavity, magnon, and mechanical modes, respectively with zero mean and characterized by correlation functions~\cite{gardinerQuantumNoiseHandbook2004}: ${\langle}c^{in}(t)c^{in\dag}(\acute{t})\rangle=[N_c(\omega_c)+1]\delta(t-\acute{t})$,  ${\langle}c^{in\dag}(t)c^{in}(\acute{t})\rangle=N_c(\omega_c)\delta(t-\acute{t})$, and ${\langle}m^{in}(t)m^{in\dag}(\acute{t})\rangle=[N_m(\omega_m)+1]\delta(t-\acute{t})$, ${\langle}m^{in\dag}(t)m^{in}(\acute{t})\rangle=N_m(\omega_m)\delta(t-\acute{t})$, and ${\langle}\xi(t)\xi(\acute{t})+\xi(\acute{t})\xi(t)\rangle/2\simeq\gamma_b[2N_b(\omega_b)+1]\delta(t-\acute{t})$, where a Markovian approximation has been made, which is valid for a large mechanical quality factor $Q=\frac{\omega_b}{\gamma_b}{\gg}1$~\cite{PhysRevA.63.023812,benguria1981quantum}. $N_j(\omega_j)=[\exp{({\hbar\omega_j}/{k_BT})}-1]^{-1}\quad (j=c,m,b)$ are equilibrium mean number of thermal photon, magnon, and phonon, respectively.

We can linearize the dynamics of the system if we strongly drive the magnon mode such that $\left|{\langle}m\rangle\right|\gg 1$.
Similarly, a strong cavity-magnon interaction leads to large cavity field amplitude $\left(\left|{\langle}c\rangle\right|\gg1\right)$.
Therefore, we can write an operator in terms of its steady-state value and a fluctuation part upto first order: $O={\langle}O\rangle+{\delta}O\quad (O=c,m,q,p)$, where ${\langle}O\rangle$ is steady-state mean value and ${\delta}O$ is a small fluctuation operator with zero mean value.
Inserting, the operator forms in Eq.~(\ref{E2}), we obtain a set of steady-state values which are nonlinear algebraic equations and a set of equations for fluctuation operators.
%The steady-state values are given by:
We write the fluctuation operator in the form of quadrature fluctuations:
\begin{align}
f(t)^T=[{\delta}x_1(t),{\delta}y_1(t),{\delta}x_2(t),{\delta}y_2(t),{\delta}q,{\delta}p],
\end{align}
with ${\delta}x_1=({\delta}c+{\delta}c^{\dag})/{\sqrt{2}}$, ${\delta}y_1=i({\delta}c^{\dag}-{\delta}c)/{\sqrt{2}}$, ${\delta}x_2=({\delta}m+{\delta}m^{\dag})/{\sqrt{2}}$, and ${\delta}y_2=i({\delta}m^{\dag}-{\delta}m)/{\sqrt{2}}$.
The resulting set of equations for quadrature fluctuations are given by:
\begin{align}\label{E3}
\dot{f(t)}=Af(t)+\sigma(t),
\end{align}
where
\begin{widetext}

\onecolumngrid
\begin{align}
\sigma(t)^T=[\sqrt{2\kappa_c}x_1^{in}(t),\sqrt{2\kappa_c}y_1^{in}(t),\sqrt{2\kappa_m}x_2^{in}(t),\sqrt{2\kappa_c}y_2^{in}(t),0,\xi(t)],
\end{align}
is the noise sources column vector, and $A$ is the drift matrix which is given by:
\begin{align}\label{E4}
A=\begin{pmatrix}
-\kappa_c+2G\cos{\theta}&\Delta_c+2G\sin{\theta}&0&g_{mc}&0&0\\-\Delta_c+2G\sin{\theta}&-\kappa_c-2G\cos{\theta}&-g_{mc}&0&0&0\\0&g_{mc}&-\kappa_m&\tilde{\Delta}_m&-G_{mb}&0\\-g_{mc}&0&-\tilde{\Delta}_m&-\kappa_m&0&0\\0&0&0&0&0&\omega_b\\0&0&0&G_{mb}&-\omega_b&-\gamma_b
\end{pmatrix}.
\end{align}
\end{widetext}
\twocolumngrid
The parameter $\tilde{\Delta}_m=\Delta_m+g_{mb}{\langle}q\rangle$ is effective magnon-drive detuning including the frequency shift due to the magnomechanical interaction, where ${\langle}q\rangle=-(g_{mb}/{\omega_b})\left|{\langle}m\rangle\right|^2$ is the steady-state position quadrature value.
The effective magnomechanical coupling rate is $G_{mb}=i\sqrt{2}g_{mb}{\langle}m\rangle$, where the steady-state value of magnon amplitude ${\langle}m\rangle$ read:
\begin{align}\label{E5}
{\langle}m\rangle={\frac{\Omega[(i\Delta_c+\kappa_c)-2Ge^{i\theta}]}{g_{mc}^2+(i\tilde{\Delta}_m+\kappa_m)[(i\Delta_c+\kappa_c)-2Ge^{i\theta}]}}.
\end{align}
The drift matrix $A$ (Eq.~(\ref{E4})) is written under the conditions:  $\left|{\tilde{\Delta}_m}\right|,\,\left|\Delta_c\right|\gg\kappa_c,\,\kappa_m, G$.
Since $\tilde{\Delta}_m$ contains $\left|{\langle}m\rangle\right|^2$, Eq.~(\ref{E5}) is intrinsically nonlinear. 
To reach a unique steady-state solution, the system should be stable.
The stability of the system is ensured by enforcing the Routh-Hurwitz criteria~\cite{PhysRevA.35.5288}, which states that the system is stable if the real parts of  all the eigenvalues of the drift matrix are negative.
We have ensured this condition in our numerical analysis.

Since, the nature of the quantum noises is assumed to be Gaussian with zero mean and we are dealing with the linearized dynamics of the quantum fluctuations, the steady-state of the quantum fluctuations is a continuous variable tripartite Gaussian state.
This state is fully characterized by a $6\times6$ covariance matrix (CM) $\mathcal{C}$, whose entries are defined as $\mathcal{C}_{ij}=\frac{1}{2}{\langle}f_i(t)f_j(\acute{t})+f_j(\acute{t})f_i(t)\rangle\,(i,j=1,2,...,6)$. 
The solution of the following Lyapunov equation (for a stable system) then gives the covariance matrix~\cite{vitali2007optomechanical}:
\begin{align}\label{E7}
A\mathcal{C}+\mathcal{C}A^T=-D,
\end{align} 
with diffusion matrix $D=diag[\kappa_c(2N_c+1),\kappa_c(2N_c+1),\kappa_m(2N_m+1),\kappa_m(2N_m+1),0,\gamma_b(2N_b+1)]$ defined through ${\langle}\sigma_i(t)\sigma_j(\acute{t})+\sigma_j(\acute{t})\sigma_i(t)\rangle/2=D_{ij}\delta(t-\acute{t})$.
Analytical solution of the Lyapunov equation for covariance matrix is cumbersome, therefore, we solve it numerically.
To quantify entanglement between any two modes of the system, we use logarithmic negativity $E_N$~\cite{li2018magnon,yang2020entanglement,PhysRevA.65.032314,PhysRevLett.95.090503}:
\begin{align}
    E_N = \max{\left[0, -\ln{2\tilde{\eta}}\right]},
    \label{EN}
\end{align}
with $\tilde{\eta}$ the minimum symplectic eigenvalue of the reduced covariance matrix of the two sub-systems of interest~\cite{li2018magnon}.
For tripartite entanglement, we use minimal residual contangle given by~\cite{li2018magnon,yang2020entanglement,Adesso_2006,Adesso_2007}:
\begin{align}\label{E9}
R_{\tau}^{min}=min[R_{\tau}^{c|mb},R_{\tau}^{m|cb},R_{\tau}^{b|cm}],
\end{align}
with residual contangle given by:
\begin{align}\label{E8}
R_{\tau}^{i|jk}=C_{i|jk}-C_{i|j}-C_{i|k}\ge 0.
\end{align}
The residual contangle is a continuous variable analogous of tangle used for discrete-variable tripartite entanglement~\cite{PhysRevA.61.052306}.
In Eq.~(\ref{E8}), the notation $C_{u|v}$ represents the contangle of sub-systems $u$ and $v$ ($v$ contains one or two modes).
For the presence of genuine tripartite entanglement in the system, there must be nonzero minimum residual contangle $R_{\tau}^{min}>0$.
Further detail on calculation of $E_N$ and $R_{\tau}$ is given in the supplementary material of Ref.~\cite{li2018magnon}.
\begin{figure}[htb] 
	\includegraphics[width=\linewidth]{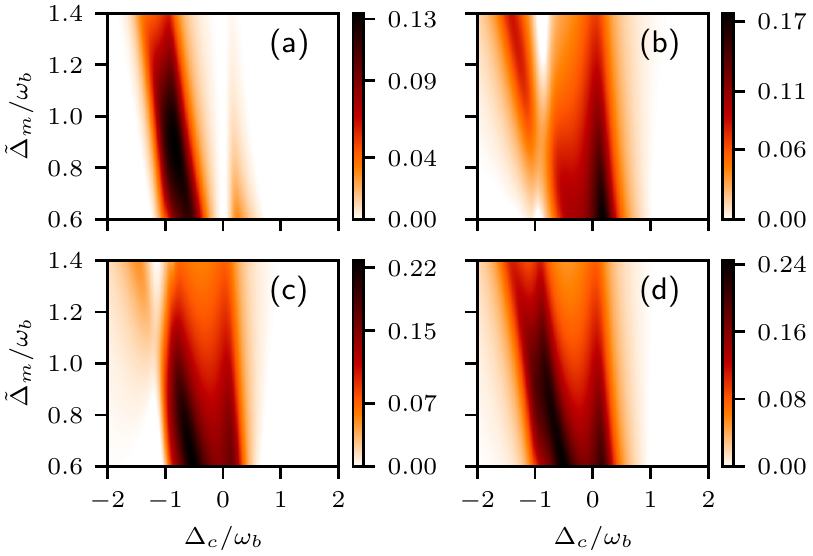}
	\caption{Density plot of bipartite entanglement between cavity and magnon modes $(E_{cm})$ versus dimensionless detunings $\Delta_c/{\omega_b}$ and $\tilde{\Delta}_m/ {\omega_b}$ (a) $G=0$, (b) - (d) $G=3.31\times10^6~s^{-1}.$ The phase $\theta$ associated with OPA drive field is assumed to be (b) $0$, (c) $\pi/2$, and (d) $\pi$, respectively.}\label{fig:2}
\end{figure}
\section{Results and Discussion}
We now present results of our numerical simulations and show how the presence of OPA not only enhances entanglement but also supports existence of entanglement in a comparatively broader range of parameter space.
This same system without OPA was recently considered by Jie Li \textit{et al.}~\cite{li2018magnon}.
Therefore, we mostly use the same experimental parameters which they considered and also reproduce some of their results for a comparison.
We adopt the following parameters~\cite{zhang2016cavity}: $\omega_c/{2\pi}=\omega_m/{2\pi}=10\,$GHz, $\omega_b/{2\pi}=10\,$MHz, $\gamma_b/{2\pi}=10^2\,$Hz, $\kappa_c/{2\pi}=\kappa_m/{2\pi}=1\,$MHz, $g_{mc}/{2\pi}=G_{mb}/{2\pi}=3.2\,$MHz, and temperature $T=10\,$mK.
Once, we choose these parameters, we found that the maximum value of $G$ which satisfies the stability criteria is  $G=3.31\times10^{6}~s^{-1}$.
We noticed that the value of phase $\theta$ associated with OPA drive field also plays an important role on the dynamics of the system, therefore, we choose three different values of $\theta$ in our numerical simulations.
\begin{figure}[htb] 
	\includegraphics[width=\linewidth]{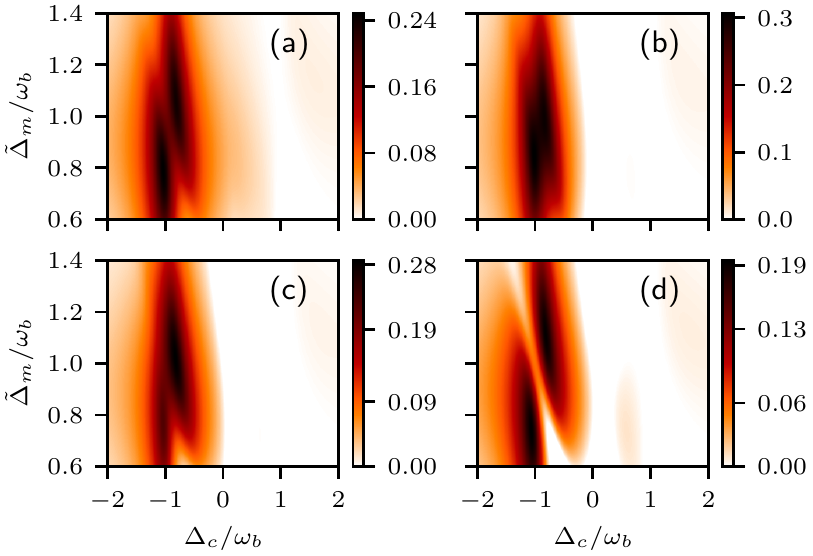}
	\caption{Density plot of bipartite entanglement between cavity and phonon modes $(E_{cb})$ versus dimensionless detunings $\Delta_c/{\omega_b}$ and $\tilde{\Delta}_m/ {\omega_b}$ (a) $G=0$, (b)-(d) $G=3.31\times10^6~s^{-1}.$ The phase $\theta$ associated with OPA drive field is assumed to be (b) $0$, (c) $\pi/2$, and (d) $\pi$, respectively.}\label{fig:3}
\end{figure}

In Fig.~(\ref{fig:2}), we present the bipartite entanglement $E_{cm}$ between the cavity and magnon mode as a function of dimensionless detunings $\Delta_c/{\omega_b}$ and $\tilde{\Delta}_m/ {\omega_b}$ in the presence as well as absence of OPA.
We reproduce the entanglement result when OPA is not present~\cite{li2018magnon} in Fig.~\ref{fig:2}(a).
In the presence of OPA, $E_{cm}$ enhances significantly as shown in Figs.~\ref{fig:2}(b) - (d).
We note that the phase of the drive field plays a crucial role as shown in Figs.~\ref{fig:2}(b) - (d), where the maximum enhancement is obtained for $\theta = \pi$.
While entanglement is enhanced roughly two times in the presence of OPA (See Fig.~\ref{fig:2}(d)), the parameter space where entanglement survives is also significantly increased.
This is an important feature which makes the experimental generation and detection of entanglement more feasible.
The enhancement in entanglement is due to the fact that the presence of OPA enhances the cavity occupation number and changes the photon statistics of the cavity mode thereby improving cavity-magnon coupling.
As a result, the steady-state value of the magnon mode (See Eq.~(\ref{E5})) also depends on parametric gain $G$ and associated phase $\theta$. 
Therefore, the phase associated with the drive field also plays an important role.
A proper choice of phase may lead to a maximum noise suppression thereby resulting maximum entanglement.
It is important to note that the effects of OPA are previously considered in cavity optomechanics showing enhancement in entanglement~\cite{Shahidani2014, Li_2015, Ahmed_2017} and cooling~\cite{Huang2009}.
\begin{figure}[htb] 
	\includegraphics[width=\linewidth]{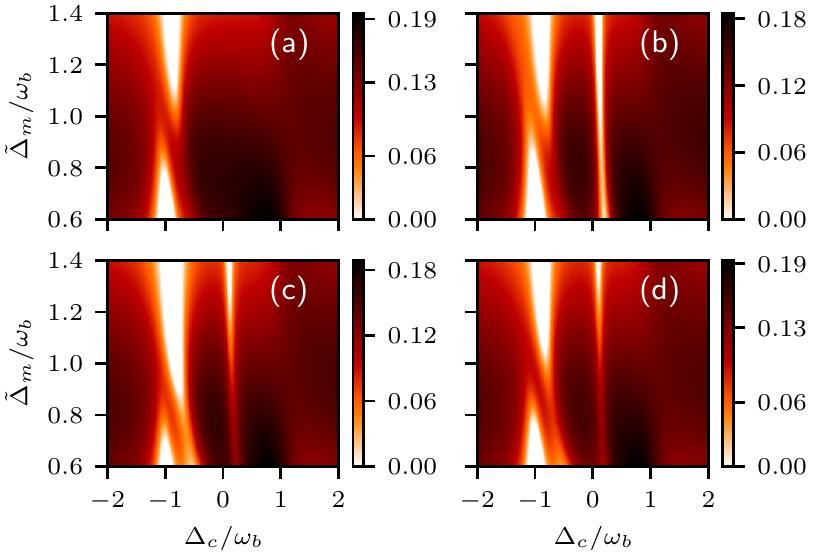}
	\caption{Density plot of bipartite entanglement between phonon and magnon modes $(E_{mb})$ versus dimensionless detunings $\Delta_c/{\omega_b}$ and $\tilde{\Delta}_m/ {\omega_b}$ (a) $G=0$, (b)-(d) $G=3.31\times10^6~s^{-1}.$ The phase $\theta$ associated with OPA drive field is assumed to be (b) $0$, (c) $\pi/2$, and (d) $\pi$, respectively.}\label{fig:4}
\end{figure}

Next, we show the effects of OPA on the bipartite entanglement between the photons and phonons $E_{cb}$.
Again, we plot $E_{cb}$ in the absence of OPA in Fig.~\ref{fig:3}(a).
The presence of the OPA enhances entanglement significantly.
The maximum enhancement occurs when the phase of the drive field is zero.
In Fig.~\ref{fig:4}, we present the effects of OPA on the bipartite entanglement between the magnon and phonon modes $E_{mb}$ of the YIG sphere.
Because the magnon mode and phonon mode are directly coupled within the YIG sphere, we do not see any appreciable effect on $E_{mb}$ due to the presence of OPA.
\begin{figure}[htb] 
	\includegraphics[width=\linewidth]{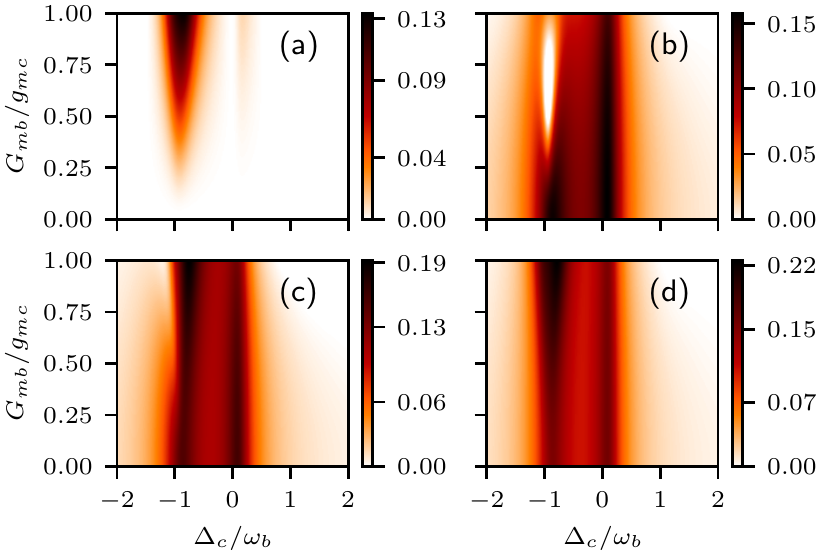}
	\caption{Density plot of bipartite entanglement between cavity and magnon modes $(E_{cm})$ versus dimensionless detuning $\Delta_c/{\omega_b}$ and dimensionless magnon-phonon coupling $G_{mb}/ g_{mc}$. (a) $G=0$, (b)-(d) $G=3.31\times10^6 s^{-1}.$ The phase $\theta$ associated with OPA drive field is (b) $0$, (c) $\pi/2$, and (d) $\pi$. In all cases, we chose $\tilde{\Delta}_m=0.9\omega_b$.}\label{fig:5}
\end{figure}

In Fig.~\ref{fig:5}, we show the density plots of bipartite cavity-magnon entanglement $E_{cm}$ versus $\Delta_c$ and $G_{mb}$ at fixed value of $\tilde{\Delta}_m=0.9\omega_b$.
In the absence of an OPA (i.e, $G=0$) and magnon-phonon coupling, cavity  and magnon modes interact via linear cavity-magnon beam
splitter interaction. 
It is well know that such interaction does not lead to entanglement~\cite{li2018magnon}.
Therefore, cavity-magnon entanglement exists only if there is a finite coupling between the magnons and phonons (a non-linear magnetostrictive interaction) [See Fig.~\ref{fig:5}(a)].
The maximum entanglement exists only if this coupling is comparable to the magnon-cavity coupling i.e, $G_{mb}\approx g_{mc}$.
Remarkably, this requirement is no longer valid if OPA is present in the cavity.
Instead we found that due to the presence of OPA, even stronger entanglement is present when $G_{mb}=0$ [See Figs.~\ref{fig:5}(b)-(d)] as compared to the $G_{mb}\approx g_{mc}$ case when there was no OPA.
Figs.~\ref{fig:5}(b)-(d) show that entanglement enhances for all the three choices of OPA phase with a maximum enhancement at $\theta=\pi$.
Moreover, entanglement is now present for a significantly wider range of $\Delta_c$, which is an important improvement from experimental point of view.
The presence of OPA changes the photon statistics by generating squeezed states of light, which then translate to magnon mode. 
The presence of squeezed states and non-linear gain of OPA generate the cavity-magnon entanglement even when magnon-phonon coupling is absent, which is quite interesting.
\begin{figure}[htb] 
	\includegraphics[width=\linewidth]{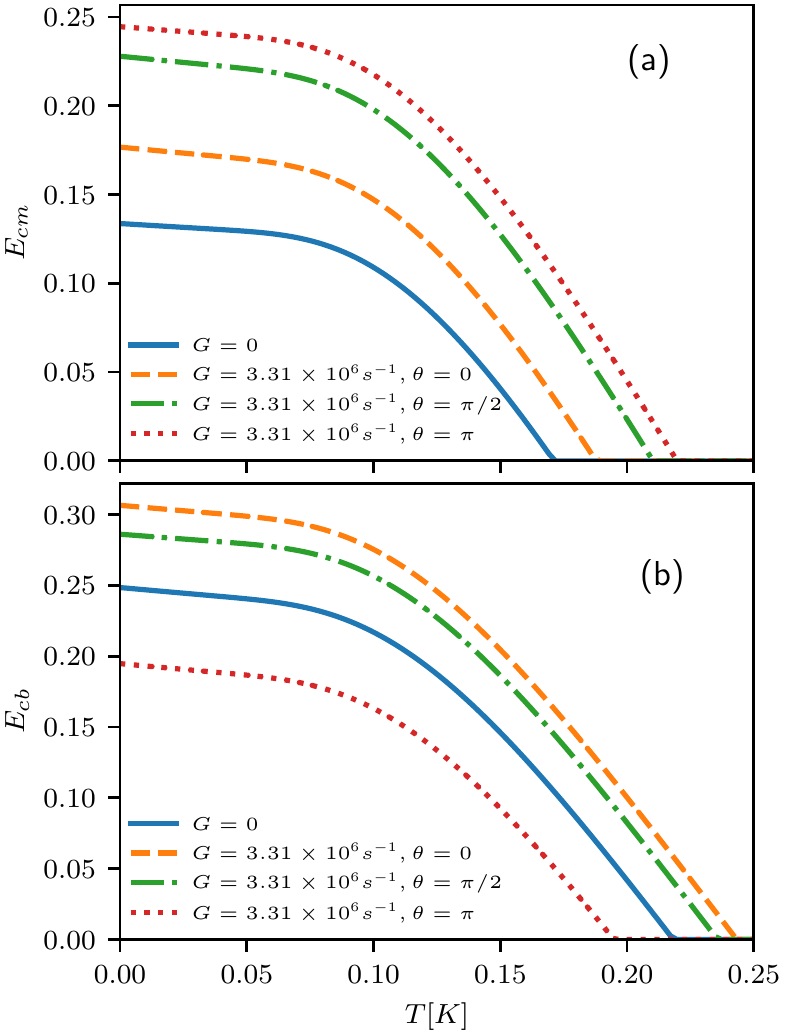}
	\caption{(a) Cavity-magnon and (b) Cavity-phonon entanglement as a function of temperature in the presence ($G=3.31\times10^6 s^{-1}$) and absence ($G=0$) of OPA. The values of $\Delta_c/{\omega_b}$ and $\tilde{\Delta}_m/ {\omega_b}$ for each curve is optimized from Fig.~\ref{fig:2} and Fig.~\ref{fig:3}.}\label{fig:6}
\end{figure}

Interestingly, the presence of OPA also enhances the robustness of entanglement against temperature as shown in Fig.~\ref{fig:6}.
In Fig.~\ref{fig:6}(a), we plot cavity-magnon entanglement as a function of temperature in the presence and absence of OPA.
We chose the optimum values of detunings $\Delta_c$ and $\tilde{\Delta}_m$ from Fig.~\ref{fig:2}.
Therefore, each curve has different optimum values corresponding to the respective panel in Fig.~\ref{fig:2}.
In the presence of OPA, the robustness of entanglement against temperature enhances significantly.
The solid curve in Fig.~\ref{fig:6}(a) shows that entanglement vanishes ($E_{cm}=0$) at a temperature of 0.17 K in the absence of OPA.
At this same temperature, the dotted curve which is plotted for $\theta=\pi$ in the presence of OPA, shows an entanglement value of $E_{cm}\approx0.11$ which is comparable to the non-OPA entanglement value of $E_{cm}\approx0.13$ at $T\approx0$.
Therefore, the presence of OPA facilitates strong entanglement at a much higher temperature.
This illustrates the importance of OPA for a robust cavity-magnon entanglement.
The dashed-dotted curve for $\theta=\pi/2$ also shows a similar behaviour where $E_{cm}\approx0.08$ at 0.17 K.
The dashed curve for $\theta=0$ also shows robustness but it is relatively less pronounced.
For cavity-phonon entanglement (See Fig.~\ref{fig:6}(b)), we found that for $\theta=0$ (dashed curve) and $\theta=\pi/2$ (dashed-dotted curve), entanglement is slightly more robust as compared to the case when OPA is not present (solid curve).
The robustness decreases when the phase $\theta=\pi$.
This is also evident from the fact that entanglement at $T\approx0$ for this case is already smaller than the entanglement when no OPA is present.

\begin{figure}[t] 
	\includegraphics[width=\linewidth]{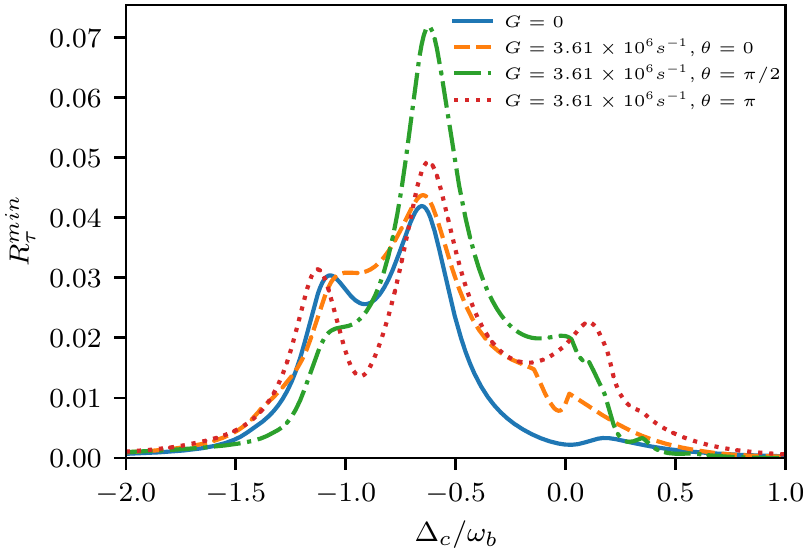}
	\caption{Tripartite entanglement between cavity-magnon-phonon modes as a function of $\Delta_c/{\omega_b}$. The solid curve is for $G=0$, while rest of the curves are for $G=3.61\times10^6 s^{-1}$ and three different choices of phase $\theta$. The value of $G_{mb}/2\pi = 4.80 MHz$. The rest of the parameters are the same as in Fig.~(\ref{fig:5}).}\label{fig:7}
\end{figure}

One of the most important feature of cavity magnomechanics is the possibility of generation of genuine tripartite entanglement in this system.
Li et al.,~\cite{li2018magnon} showed that genuine tripartite magnon-phonon-photon entanglement exists in the system, if the magnon mode is in resonance with anti-Stokes (blue sideband) and the cavity mode is in resonance with Stokes (red sideband).
Next, we consider the effects of OPA on tripartite entanglement.
In order to study this, we plot the minimum of the residual contangle in Fig.~\ref{fig:7} as a function of detuning $\Delta_c$.
Here, we use $G_{mb}/2\pi = 4.8~MHz$ \cite{li2018magnon}.
The solid curve represents the tripartite entanglement in the absence of OPA \cite{li2018magnon}.
For a non-zero value of gain parameter $G=3.61\times10^6 s^{-1}$, we plot $R_{\tau}^{min}$ for $\theta=0$, $\pi/2$, and $\pi$ in dashed, dashed-dotted, and dotted curve, respectively.
For $\theta=\pi/2$, we obtained maximum enhancement in tripartite entanglement.
The enhancement is also accompanied with an increase in domain of entanglement over a relatively wider space of detuning values.
For other two choices, there is no significant enhancement in the maximum value, however, the overall area in which tripartite entanglement exists increases which is quite interesting.

It is important to note that we have neglected Kerr effect in our study.
The presence of the Kerr effect has very important consequences as discussed in Ref.~\cite{yang2020entanglement, Hyde-18}.
In fact, Kerr effect strongly depends on the volume of the sphere and the strength of drive field.
For a YIG sphere of diameter $d=250\mu m$ and using the other relevant parameters considered here, we show that such nonlinear effects are negligible and can be ignored in our linearized model. 
Since $\left|{\tilde{\Delta}_m}\right|,\,\left|\Delta_c\right|\gg\kappa_c,\,\kappa_m, G$ and $ g_{mc}^2 \ll \left|{\tilde{\Delta}_m \Delta_c}\right|\approx \omega_b^2$ for the parameters used in our simulations, the effective magnon-phonon coupling is given by $G_{mb} = \sqrt{2}g_{mb} \Omega/\omega_b$~\cite{li2018magnon}.
For $G_{mb}/2 \pi = 3.2 MHz$ and $g_{mb} /2 \pi = 0.2 Hz$, the strength of the magnetic field is approximated $B_0\approx 3.9\times 10^{-5}T$.
The number of spins in the YIG sphere considered here are $N\approx3.5 \times 10^{16}$ and therefore, the corresponding drive strength $\Omega \approx 7.1\times 10^{14}Hz$.
In an earlier study Li et. al., \cite{li2018magnon} estimated that the magnitude of Kerr coefficient for a $250\mu m$ sphere is $K\approx 6.4nHz\ll \omega_b^3/\Omega^2$, which is negligible and our linearized model is valid for the considered sphere size.
Nevertheless, including both Kerr effect and OPA may lead to interesting new results and may be considered in a future study.

\section{Conclusion}
In conclusion, we studied a scheme to enhance bipartite and tripartite entanglement in cavity magnomechanical system.
We found that the presence of OPA in cavity magnomechanical system significantly enhances the bipartite cavity-magnon and cavity-phonon entanglement for a proper choice of OPA phase.
Interestingly, the presence of OPA has no significant effect on magnon-phonon entanglement.
This is due to the fact that the magnon and phonon modes are directly coupled within YIG sphere and OPA only changes the photon statistics.
Another important finding is that the presence of OPA relaxes the condition of strong magnon-phonon coupling to achieve cavity-magnon entanglement.
We found strong cavity-magnon entanglement even in the absence of magnon-phonon coupling.
The robustness of both cavity-magnon and cavity phonon entanglement against temperature also enhances in the presence of OPA.
Most importantly, the OPA enhances the tripartite entanglement significantly in the limit of experimentally feasible parameters.
Apart from the above mentioned enhancements, OPA also increases the domain of entanglement in the parameter regime which is an important feature related to the possible experimental implementation.
\bibliography{./document}
\end{document}